\documentclass[aps,reprint, prl, superscriptaddress, showpacs,nobibnotes]{revtex4-1}
\usepackage{graphicx}
\usepackage{bm,amsmath}
\usepackage{dcolumn}
\usepackage{natbib,xcolor}

\begin{document}
\title{Field-dependent Shubnikov-de Haas oscillations in ferromagnetic Weyl semimetal Co$_3$Sn$_2$S$_2$}

\author{Linda Ye}
\thanks{These authors contributed equally}
\homepage{Present Address: Department of Applied Physics, Stanford University, Stanford, California 94305, USA}
\affiliation{Department of Physics, Massachusetts Institute of Technology, Cambridge, Massachusetts 02139, USA}
\author{Jorge I. Facio}
\thanks{These authors contributed equally}
\homepage{Present Address: Centro At\'omico Bariloche, Instituto de Nanociencia y Nanotecnolog\'ia (CNEA-CONICET) and Instituto Balseiro. Av. Bustillo 9500, Bariloche (8400), Argentina. }
\affiliation{Leibniz IFW Dresden, Helmholtzstraße 20, 01069 Dresden, Germany}
\author{Madhav P. Ghimire}
\affiliation{Central Department of Physics, Tribhuvan University, Kirtipur 44613, Kathmandu, Nepal}
\affiliation{Leibniz IFW Dresden, Helmholtzstraße 20, 01069 Dresden, Germany}
\author{Mun K. Chan}
\affiliation{National High Magnetic Field Laboratory, LANL, Los Alamos, New Mexico 87545, USA}
\author{Jhih-Shih You}
\affiliation{Department of Physics, National Taiwan Normal University, Taipei 11677, Taiwan}
\author{David C. Bell}
\affiliation{Harvard John A. Paulson School of Engineering and Applied Sciences, Harvard University, Cambridge, Massachusetts 02138, USA}
\affiliation{Center for Nanoscale systems, Harvard University, Cambridge, Massachusetts 02138, USA}
\author{Manuel Richter}
\affiliation{Leibniz IFW Dresden, Helmholtzstraße 20, 01069 Dresden, Germany}
\affiliation{Dresden Center for Computational Materials Science (DCMS), TU Dresden, 01062 Dresden, Germany}
\author{Jeroen van den Brink}
\affiliation{Leibniz IFW Dresden, Helmholtzstraße 20, 01069 Dresden, Germany}
\affiliation{W\"urzburg-Dresden Cluster of Excellence ct.qmat, Technische Universit\"at Dresden, 01062 Dresden, Germany}
\author{Joseph G. Checkelsky}
\affiliation{Department of Physics, Massachusetts Institute of Technology, Cambridge, Massachusetts 02139, USA}

\date{\today}
\begin{abstract}
We report a study of Shubnikov-de Haas oscillations in high quality single crystals of ferromagnetic Weyl semimetal Co$_3$Sn$_2$S$_2$. The Fermi surfaces resolved in our experiments are three-dimensional and reflect an underlying trigonal crystallographic symmetry. Combined with density functional theoretical calculations, we identify that the majority of the Fermi surfaces in the system -- of both electron and hole nature -- arise from the strong energy dispersion of the (spin-orbit gapped) mirror-protected nodal rings. We observe that an in-plane magnetic field induces a continuous evolution of Fermi surfaces, in contrast to field perpendicular to the kagome lattice planes which has little effect. Viewed alongside the easy-axis anisotropy of the system, our observation reveals an evolution of the electronic structure of Co$_3$Sn$_2$S$_2$ -- including the Weyl points -- with the ferromagnetic moment orientation. Through the case study of Co$_3$Sn$_2$S$_2$, our results provide concrete experimental evidence of an anisotropic interplay via spin-orbit coupling between the magnetic degrees of freedom and electronic band singularities, which has long been expected in semimetallic and metallic magnetic topological systems.
\end{abstract}
\maketitle

Topological semimetals (TSM) refer to a class of bulk gapless topological phases that host singular electronic excitations in three-dimensional momentum space \cite{WSM}. A particularly versatile subclass of TSMs are those who in the meantime possess magnetic order \cite{MSM_calculation,Magnetic_database}; in such magnetic TSMs, coexisting band topology and magnetic order parameters in principle allows the manipulation of the former using the latter. Early materials proposals of magnetic TSMs have built heavily on insights gained from the non-magnetic sector -- pioneering ideas include modulation-doped topological insulating heterolayers \cite{MTI_layer} and stoichiometric magnetic materials whose non-magnetic analogues are known as host to topologically non-trivial bands \cite{HgCr2Se4,GdPtBi}. In those systems band topology and magnetism are typically carried by separate orbitals from distinct elements: itinerant $p$ orbitals often form topological bands while localized $d$ and $f$ electrons support magnetism \cite{HgCr2Se4,GdPtBi,RAlGe}. A particularly exciting recent advancement is the discovery of a class of systems where the correlated 3$d$-electrons play the dual role of driving magnetic orders and composing topological bands \cite{Co2MnGa_ANE,Mn3Ge,Mn3Sn_Weyl,Fe3Sn2,Co3Sn2S2_AHE,wang2018large}. This not only enables studies of topological responses of magnetic TSMs beyond room temperature \cite{Co2MnGa_ANE,Mn3Ge,Mn3Sn_Weyl,Fe3Sn2}, but also in principle dictates an intrinsic coupling between magnetic order and electronic topology. 

In this Letter, via a combined Shubnikov-de Haas (SdH) oscillation and density functional theory (DFT) study, we explore the intertwined ferromagnetic order and the Weyl semimetallic phase in Co$_3$Sn$_2$S$_2$.
The emergence of Weyl fermions in Co$_3$Sn$_2$S$_2$ \cite{Co3Sn2S2_AHE,wang2018large,Co3Sn2S2_Weyl_ARPES,morali2019fermi} can be viewed in a similar manner with inversion-symmetry-breaking non-magnetic Weyl semimetals \cite{TaAs_theory,TaAs1,TaAs2} where an inverted pair of bands first cross and generate nodal rings; then the spin-orbit coupling (SOC) in the presence of ferromagnetism gaps out the ring except at isolated points, the Weyl nodes. 
Here via high field fermiology studies on high quality single crystals, we experimentally demonstrate the evolution of the electronic structure with the orientation of the ferromagnetic moments in the system. With DFT we show that the observed evolution manifests an intrinsic connection between magnetism and topology:  a strong dependence of the SOC-induced gap along the nodal ring with the magnetic moment orientation.

\begin{figure*}[t]
	\includegraphics[width = 2\columnwidth]{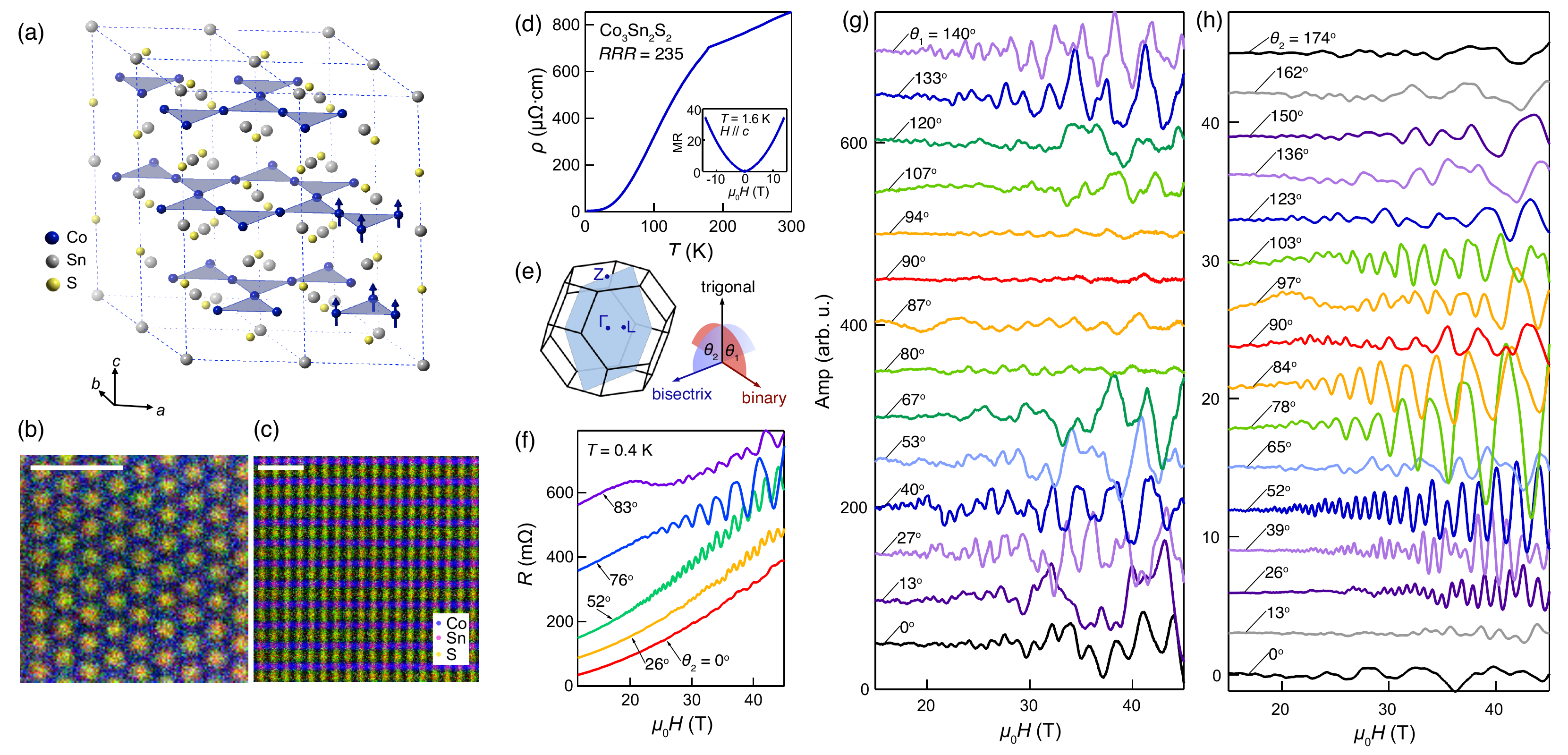}
	\caption{\label{fig-1} (a) Crystal structure of Co$_3$Sn$_2$S$_2$ with Co atoms shown in blue, Sn in gray, and S in yellow. The blue arrows show the Co ferromagnetic moments. (b,c) Transmission electron microscope image of Co$_3$Sn$_2$S$_2$ viewed from the $c$-axis (b) and the $a^*$-axis (c), respectively. In (b,c) Co intensities are shown in blue, Sn in magenta, and S in yellow. The scale bars represent 1 nm. (d) Resistivity as a function of temperature $T$ of a typical CVT Co$_3$Sn$_2$S$_2$ single crystal. The inset shows the magnetoresistance (MR) defined as $(\rho(T,H)-\rho(T,0))/\rho(T,0)$ taken at $T=1.6$ K and with field applied along $c$. (e) Schematic of magnetic field rotation shown along with the rhombohedral Brillouin zone of Co$_3$Sn$_2$S$_2$, where $\theta_1$ is defined within the trigonal ($c$-axis)-bisetrix plane, $\theta_2$ within the trigonal-binary plane. (f) Resistance as a function of field up to 45 T at 0.4 K in a $\theta_2$ rotation. (g,h) Oscillatory part of the resistance in the magnetic field at selected angles in a $\theta_1$ rotation (g) and $\theta_2$ rotation (h), respectively.}
\end{figure*}

\textit{Quantum oscillations and Fermi surfaces of Co$_3$Sn$_2$S$_2$}
Co$_3$Sn$_2$S$_2$ crystallizes in a Shandite Ni$_3$Pb$_2$S$_2$-type structure (Fig. \ref{fig-1}(a));  the Co atoms form kagome lattices that are A-B-C stacked along the $c$-axis. Co$_3$Sn kagome layers spaced by additional Sn and S atoms can be identified in element-resolved Transmission Electron Microscopy images of our single crystals in Fig. 1(b,c). We show in Fig. \ref{fig-1}(d) the resistivity $\rho$ of a typical CVT crystal as a function of temperature $T$. The residual resistivity ratio (RRR) defined as $\rho(T = 300$\,K) $/\rho(T = 2$\,K) shows an enhanced value 235 and a low residual resistivity 3.5 $\mu\Omega$cm compared to previous reports \cite{Co3Sn2S2_AHE,Co3Sn2S2_Behnia}. The magnetoresistance at $T=1.6$\,K is shown in Fig. \ref{fig-1}(d) inset where Shubnikov-de Haas oscillations are resolved from 5\,T (see Supplementary Materials \footnote{See the Supplementary Materials, which includes citations to Refs. \cite{PhysRevB.59.1743,lejaeghere2016reproducibility,varjas2018qsymm}, for data on on quantum oscillations for different temperatures, FFT spectra, methodological aspects of the DFT calculations and further details on the $\mathbf{k}\cdot\mathbf{p}$ model.}). Together with the Hall traces, using a two band model we can fit $\rho_{xx}$ and $\rho_{yx}$ simultaneously with $n_1=9.5\times 10^{19}/\text{cm}^3, \mu_1=6.3\times 10^{3}\mathrm{cm}^2/\text{V}\cdot \text{s}, n_2=-8.1\times 10^{19}/\text{cm}^3, \mu_2=7.3\times 10^{3}\text{cm}^2/\text{V}\cdot \text{s}$ \cite{Note1}. The high electronic quality in our CVT crystals allows us to experimentally study of the Fermi surfaces of the system via magneto-quantum oscillations.

\begin{figure*}
	\includegraphics[width =2\columnwidth]{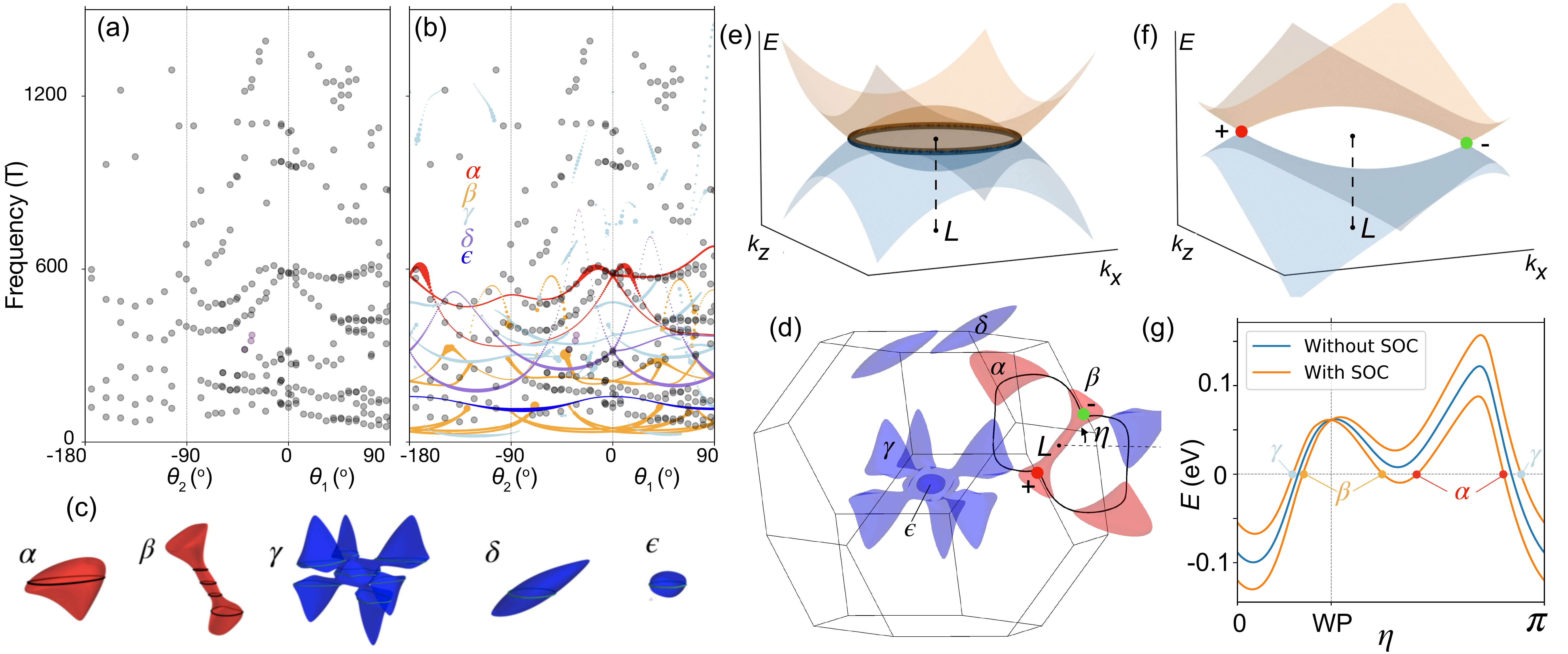}
	\caption{(a) Experimental quantum oscillation frequencies as a function of the orientation of applied field $H$. (b) Comparison between the theoretical quantum oscillation spectra derived from DFT (colored lines) and experimental points (gray circles). (c,d) DFT Fermi surfaces of Co$_3$Sn$_2$S$_2$ $\alpha,\beta,\gamma,\delta$ and $\epsilon$ shown along with extremal cyclotron orbits for an out-of-plane magnetic field (c) and in the BZ (d). For clarity we show in (d) a subset of symmetry-related Fermi pockets. In (c,d) the hole Fermi surfaces ($\alpha,\beta$) are shown in red and electron pockets ($\gamma,\delta, \epsilon$) in blue. (e,f) Energy dispersion at the reflection-symmetry plane $k_y=0$ for a  $\mathbf{k} \cdot \mathbf{p}$  model obeying the material symmetries for zero and finite SOC,  respectively. (g) Energy dispersion of Co$_3$Sn$_2$S$_2$ along the nodal ring; blue and orange lines are obtained without and with SOC, respectively. $\eta$ labels the angle along the nodal ring with respect to the horizontal dashed line starting from $L$ as illustrated in (d). We also mark the Fermi points corresponding to $\gamma,\beta$ and $\alpha$ pockets in (g). }
	\label{fig-fs}
\end{figure*}

In Fig. \ref{fig-1}(f) we show the magnetoresistance traces measured at the NHMFL 45\,T Hybrid magnet. The magnetic field $H$ is rotated from the trigonal to either the binary or the bisectrix axis, defining respectively $\theta_1-$ and $\theta_2-$ rotations schematically illustrated in Fig. \ref{fig-1}(e). At $T= 0.4$\,K, we show the evolution of resistance with $H$ at selected $\theta_2$ in Fig. 2(f), where prominent Shubnikov-de Haas (SdH) oscillations can be identified. When $H$ is applied close to the kagome planes (see \textit{e.g.} the trace with $\theta_2=83^{\circ}$ in Fig. \ref{fig-1}(f)), we can identify a kink in the magnetoresistance near 20\,T, which we assign to a critical field beyond which the alignment of the magnetic moments with $H$ in the hard plane is complete (this is comparable with that extrapolated from the low field magnetization \cite{Co3Sn2S2_PRR_Ghimire}). We return to the consequences of the reorientation process of the ferromagnetic moments below. We show post-background-subtraction SdH oscillations for $\theta_1$ and $\theta_2$ rotations in Figs. \ref{fig-1}(g) and (h), respectively. That we identify SdH oscillations at all angle suggests a three-dimensional nature of the underlying Fermi surfaces in the system.

In Fig. \ref{fig-fs}(a) we summarize the frequencies obtained from a Fast Fourier Transformation (FFT) of the SdH oscillations between 11.5\,-\,45\,T (FFT spectra themselves are shown in \cite{Note1}) with the orientation of $H$. We show $\theta_1$ ($\theta_2$) between $0^{\circ}\,-\,90^{\circ}$ ($0^{\circ}\,-\,180^{\circ}$) because the oscillation frequencies appear symmetric (asymmetric) with respect to $90^{\circ}$ for $\theta_1$ ($\theta_2$) \cite{Note1}, consistent with the trigonal symmetry of the underlying crystal structure and electronic structure of Co$_3$Sn$_2$S$_2$ (a result of ABC stacking of the kagome lattices). Fig. \ref{fig-fs}(b) compares the experimental results with simulated quantum oscillation spectra based on DFT Fermi surfaces. The theoretical results are shown with point sizes proportional to an amplitude reduction factor $R_DR_T/m$ introduced to simulate the relative expected oscillation amplitude across the different sets of Fermi surfaces.  Here, $R_D$ is the Dingle factor, $R_T$ the Lifshitz-Kosevich factor, which describes the disorder and temperature smearing effects, respectively; $m$ is proportional to the curvature of Fermi surface cross section along $H$ \cite{Note1}. The Fermi surfaces giving rise to oscillation branches $\alpha,\beta,\gamma,\delta$ and $\epsilon$ are highlighted individually in Fig. \ref{fig-fs}(c) and are shown in the BZ in Fig. \ref{fig-fs}(d). We obtain an overall qualitative agreement.  We highlight the comparison for the hole pocket $\alpha$-associated branches in Fig. \ref{fig-fs}(b). The observed set of frequencies $>800$ T can be captured by $\gamma$ being a single connected piece of Fermi surface as shown in \ref{fig-fs}(c); such Fermi surface topology of $\gamma$ is consistent with Refs. \cite{PhysRevB.103.205112,PhysRevLett.124.077403} but distinct from disconnected fin-like Fermi pockets reported in a number of other works \cite{Co3Sn2S2_AHE,Co3Sn2S2_Behnia}. This discrepancy among calculations can be attributed to the proximity of $E_F$ to a van Hove singularity and thus a Lifshitz transition; as a result different calculation schemes can yield abrupt changes in the connectivity of $\gamma$ near the BZ center \cite{Note1}. For low frequencies below 500\,T, a one-to-one assignment of the calculated Fermi surfaces is challenging due to the presence of multiple frequencies of comparable size. 

\textit{Nodal ring in Co$_3$Sn$_2$S$_2$.} 
We now introduce a two-band $\mathbf{k}\cdot\mathbf{p}$ model to explain how the rather complex observed Fermi surfaces connect to the underlying topological features. We start around the point $L$, the center of hexagonal facets of the BZ (see Fig. \ref{fig-fs}(d)). $L$ is invariant under inversion $I$ and reflection $M_y$ symmetries, in the following represented by $I=\sigma_z$ and $M(y)=-\sigma_z$. For finite SOC $\lambda$, the electronic symmetry depends on the direction of the magnetization, here taken to be the $c$ axis. In this case, $M_y$ is replaced by the antiunitary mirror symmetry $M_y T$, with $T=i \sigma_z K$ the time-reversal operator and $K$ the complex conjugation. We consider the effective Hamiltonian
\begin{align}
	H(\vec{k}) &= \begin{bmatrix}
0 & 0 \\
0 & \Delta 
\end{bmatrix} + \alpha k_y \sigma_x + (\beta_x k_x^2 + \beta_z k_z^2) \sigma_z \nonumber\\
	&+ \lambda (k_x+k_z) \sigma_y,
	\label{eq_model}
\end{align}
where the momenta are measured with respect to $L$. It obeys inversion symmetry and, depending on $\lambda$, it is symmetric under $M_y$ or $M_y T$ \cite{Note1}. 
For $\lambda=0$, the salient feature  is a nodal ring within the mirror-invariant $k_y=0$ plane defined by the locus $\beta_x k_x^2 + \beta_y k_y^2=\Delta/2$ (see Fig. \ref{fig-fs}c).
For finite $\lambda$, such line becomes gapped except at the two points $\mp \beta_x k_x = \pm\beta_z k_z = \sqrt{\Delta}/2$ (Fig. \ref{fig-fs}d). In Co$_3$Sn$_2$S$_2$, the pair of bands closest to $E_F$ are exactly half-filled; this can potentially lead to an ideal Weyl semimetal as we illustrate in Fig. \ref{fig-fs}(d). The complexity of the real Fermi surface originates mainly in the energy dispersion ($\sim$200 meV) of the nodal ring, which deforms the Weyl bands  from the ideal case, yielding a rich energetic landscape (Fig. \ref{fig-fs}(g)): in addition to $\beta$ that encloses the Weyl points, $\alpha$ and $\gamma$ can also be attributed to the nodal ring energy dispersion. Additional electron pockets can be found near both $\mathrm{\Gamma}$ ($\epsilon$ and a portion of $\gamma$) and along $\mathrm{L}-\mathrm{M}$ ($\delta$) \cite{Note1}; the former involves a third band. 

\begin{figure}
	\includegraphics[width =  \columnwidth]{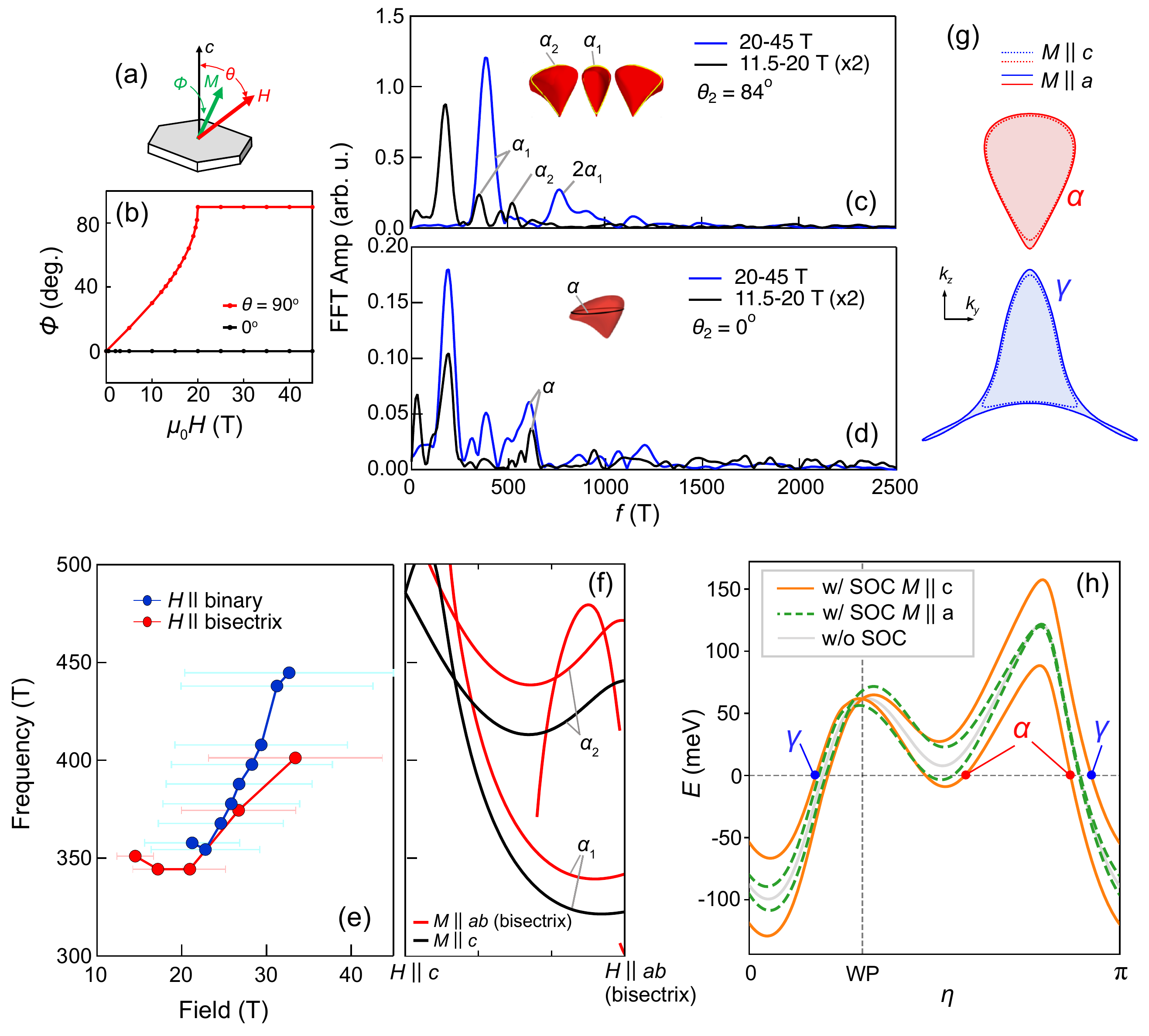}
	\caption{(a) Schematic of magnetic field $H$ and ferromagnetic moment $M$ with angles $\theta$ and $\phi$ with respect to the $c$-axis, respectively. (b) Simulated evolution of $\phi$ with $H\parallel ab$ (red curve) and $H\parallel c$ (black curve). (c,d) FFT spectra obtained with magnetic field range 11.5 T -20 T (black) and 20 T -45 T (blue) with a nearly in-plane $H$ (c) and an out-of-plane $H$ (d). (e) Evolution of FFT peak position for $\alpha_1$ with a moving FFT window; the field range is indicated by the horizontal lines. (f) Calculated oscillation frequencies with $M$ along the trigonal (black curves) or the bisectrix direction (red curves). (g) Extremal orbits of $\alpha$ and $\gamma$ viewed from the bisectrix direction. The dashed lines are obtained with $M\parallel c$ and solid lines with $M\parallel ab$. (h) Energy dispersion of the nodal ring with $M\parallel c$ (orange line) and $M\parallel a$ (green dashed line) as well as for without SOC (gray line).}\label{fig_rotation}
\end{figure}

\textit{Fermi surface changes due to rotation of the magnetization.} 
The orientation of the magnetization ($M$) has been shown to  tune the electronic topology in various ferromagnets \cite{PhysRevB.92.085138,PhysRevB.96.201102,PhysRevB.98.121103,Co3Sn2S2_PRR_Ghimire,ray2022tunable,kumar2021tuning}, including the Weyl-node energy in the case of Co$_3$Sn$_2$S$_2$ \cite{Co3Sn2S2_PRR_Ghimire}. Below we show that the observed SdH oscillations (Fig. \ref{fig-1}(g,h)) allow us to examine electronic structure reconstructions associated with a rotation of $M$. To illustrate the experimental window available in our setup, we consider a model consisting of an uniaxial magnetic anisotropy term and a Zeeman energy: $E=DM^2\sin^2\phi-MH\cos(\theta-\phi)$. Here $\theta$ ($\phi$) is the angle of $H$ ($M$) from the $c$-axis (Fig. \ref{fig_rotation}(a)) and we take $D$ such that the field required to fully align $M$ with $H\parallel ab$ is 20\,T. As Fig. \ref{fig_rotation}(b) shows, when $H$ is applied in the kagome plane, we expect a gradual and then accelerated rotation of $M$ into the plane with increasing $H$. 

In Fig. \ref{fig_rotation}(c) we show the FFT spectra between 11.5\,T - 20\,T and 20 - 45\,T, where $M$ is expected to be continuously rotating into the $ab$ plane within the former field range, and subsequently pinned along $H$ within the latter. A strong modification of the FFT spectra with the corresponding field range can be identified in Fig. \ref{fig_rotation}(c). In contrast, performing a similar analysis for $H\parallel c$ reveals an FFT spectra with peak frequencies invariant with the range of $H$ (Fig. \ref{fig_rotation}(d)), consistent with $M$ in the latter case always pointing along  the $c$-axis in quantizing fields (Fig. \ref{fig_rotation}(b)) (thus no electronic structure reconstruction is expected).
Contrasting Fig. \ref{fig_rotation}(c) and (d) suggests that the Fermi surfaces thus the underlying electronic structure are evolving with the orientation of $M$ in Co$_3$Sn$_2$S$_2$. 

We now focus on the hole pocket $\alpha$ which exhibits the strongest agreement between the observed quantum oscillation frequencies and DFT. Between the two subbranches of $\alpha$ we consider the lower $\alpha_1$ (cyclotron orbit illustrated in Fig. \ref{fig_rotation}(c) inset). In Fig. \ref{fig_rotation}(e) we show the evolution of $\alpha_1$ obtained with a moving FFT window with $H\parallel ab$. With $H$ along the bisectrix (binary) direction, we observe an increase of approximately 15\% (20\%) for $\alpha$; both suggest that $\alpha$ grows in size with an in-plane magnetic field.

In Fig. \ref{fig_rotation}(f) we compare the calculated oscillation spectra of $\alpha$ for both $M\parallel c$ (black curves) and $M\parallel ab$ (red curves). Although only a subset of ($M,H$) configurations are experimentally accessible due to the strong uniaxial anisotropy, in Fig. \ref{fig_rotation}(f) we examine the theoretical cyclotron orbits perpendicular to all orientations of $H$ to carve out the overall shape of $\alpha$. For $M\parallel ab$ there is an overall increase of the pocket size of $\alpha$ compared to $M\parallel c$; the calculated Fermi pocket size increase ($\sim10\%$) is comparable with albeit smaller than the experiments ($15\sim20\%$). We illustrate in Fig. \ref{fig_rotation}(g) the cross section of $\alpha_1$ along with that of $\gamma$ in its immediate neighborhood: an in-plane $M$ (solid lines in Fig. \ref{fig_rotation}(g)) appear to increase the size of both $\alpha$ and $\gamma$ pockets and decreases the gap between the two as compared to $M\parallel c$ (dashed lines in Fig. \ref{fig_rotation}(g)). Remarkably, the root of this phenomenology lies in a strong suppression of the energy gap along the nodal ring (Fig. \ref{fig_rotation}(h)) when $M$ rotates into the $ab$-plane. Ultimately, our observations provide experimental evidence for an anisotropic coupling between the ferromagnetic moments with the nodal ring states via SOC. 

In summary, we have identified three-dimensional Fermi surfaces of Co$_3$Sn$_2$S$_2$ in a combined SdH and DFT study. A majority of Fermi surfaces in the system are found to derive from the strong energy dispersion of the nodal rings on the (110) mirror planes. Our high field fermiology study has allowed the observation of signatures of moment-orientation induced modulation of the SOC gap along the nodal line, supporting the theoretical scenario that new Weyl points can be generated and their energy tuned by an in-plane orientation of the magnetic moments \cite{Co3Sn2S2_PRR_Ghimire}. Although Co$_3$Sn$_2$S$_2$ is widely viewed as a ferromagnetic Weyl semimetal, our study of the bulk Fermi surfaces of Co$_3$Sn$_2$S$_2$ reinforces the central role of the nodal ring in understanding its transport and optical responses, along with its electronic topology \cite{Co3Sn2S2_AHE,wang2018large,Co3Sn2S2_optical,Co3Sn2S2_ARPES_loop}. The implications of our quantum oscillation studies in clean samples of Co$_3$Sn$_2$S$_2$ can also be applied to a broad class of magnetic TSMs hosting point and line nodes (\textit{e.g.} Mn$_3$(Ge,Sn) \cite{Mn3Ge,Mn3Sn_Weyl}, Co$_2$MnGa \cite{Co2MnGa_ANE}, Fe$_3$GeTe$_2$ \cite{Fe3GeTe2}).

Viewed alternatively from the perspective of kagome lattice metals, we may contrast the 3D Fermi surfaces observed here in Co$_3$Sn$_2$S$_2$ with the quasi-two-dimensional Dirac fermiology reported in Fe$_3$Sn$_2$ \cite{Fe3Sn2,Fe3Sn2_dHvA} and FeSn \cite{FeSn}. The tunability of the dimensionality of the topological states in magnetic kagome metals may be attributed to the orbital degrees of freedom of the underlying $3d$ electrons \cite{Co3Sn2S2_Nomura,Fe3Sn2_DFT,FeSn}. With increasing interplane hopping, ferromagnetic kagome metals connect a 3D quantum anomalous Hall insulator phase to a 3D Weyl semimetallic phase, suggesting a topological phase diagram resembling the seminal theoretical proposal of building Weyl semimetallic phases using stacked topological insulating layers \cite{MTI_layer}. Additionally, rich magnetic phases have been reported for Co$_3$Sn$_2$S$_2$ at low magnetic fields \cite{Co3Sn2S2_EB,Co3Sn2S2_lowfield} as signatures of putative magnetic frustration in addition to the dominant ferromagnetic interactions; in view of the intertwined magnetic order and the topological electronic state in Co$_3$Sn$_2$S$_2$ revealed in the present study, the system provides an exciting platform to study the interplay between frustration and band topology.

\emph{Acknowledgments}
We acknowledge S. Fang, M. Kang, and R. Comin for fruitful discussions and U. Nitzsche for technical assistance. This research is funded in part by the Gordon and Betty Moore Foundation’s EPiQS Initiative, Grant GBMF9070 to J.G.C. (material synthesis), ARO Grant No. W911NF-16-1-0034 (advanced characterization), and ONR Grant No. N00014-21-1-2591 (analysis).  We also acknowledge financial support from the German Research Foundation (Deutsche Forschungsgemeinschaft, DFG) via SFB1143 Project No. A5 and under Germany’s Excellence Strategy through Würzburg-Dresden Cluster of Excellence on Complexity and Topology in Quantum Matter—ct.qmat (EXC 2147, Project No. 390858490). L.Y. acknowledges support by the Tsinghua Education Foundation and the STC Center for Integrated Quantum Materials, NSF grant number DMR-1231319.  J.I.F. and M.P.G. would like to acknowledge the support from the Alexander von Humboldt Foundation. A portion of this work was performed at the National High Magnetic Field Laboratory, which is supported by the National Science Foundation Cooperative Agreement No. DMR-1157490 and DMR-1644779, the State of Florida, and the U.S. Department of Energy. Pulsed magnetic field measurements at Los Alamos National Laboratory were supported by the U.S. Department of Energy BES “Science at 100T” grant.

\bibliography{references}
\pagebreak
\end{document}